\documentclass{scrartcl}

\usepackage{amsmath,amssymb}
\usepackage[noblocks]{authblk} 
\usepackage{hyperref}
\usepackage{multicol}
\usepackage{multirow}
\usepackage[normalem]{ulem}
\usepackage{longtable}
\usepackage{booktabs}
\usepackage{array} 
\usepackage{calc}
\usepackage{graphicx} 
\usepackage{chngcntr}
\usepackage{tocloft}
\usepackage[
style=authoryear,
natbib=true,
backend=biber
]{biblatex} 
\usepackage{tabularx} 

\newcolumntype{M}[1]{>{\centering\arraybackslash}m{#1}} 

\usepackage[left=2cm, right=2cm, top=3cm, bottom=3cm]{geometry}

\graphicspath{{images/}}

\title{Phosphorus recycling from human excreta in French agroecosystems and potential for food self-sufficiency}
\author[1]{Thomas Starck\footnote{thomas.starck@polytechnique.org}}
\author[1]{Tanguy Fardet}
\author[1,2]{Fabien Esculier}
\affil[1]{LEESU, Ecole des Ponts, Univ Paris Est Creteil, Marne-la-Vallée, France}
\affil[2]{METIS, Sorbonne Université, CNRS, EPHE, Paris, France}

\date{}

\begin{document}

\maketitle

\vspace{-4em}

\section*{Abstract}

\begin{abstract}
Phosphorus (P) is an essential constituent of life but large P losses from agroecosystems and sanitation systems are a major source of eutrophication in water bodies.
These losses are doubly negative as P in human excretions can be used for crop fertilization.
Using a unique dataset of 20,000 French WasteWater Treatment Plants (WWTPs) operational measurements over two decades and a P mass balance of the sanitation system, we assess the fate of human excretions and their agricultural potential.
Despite 75\% of French WWTPs sludge being applied to crops, only 50\% of the excreted P is returned to agroecosystems. This is among the highest rate in Western countries.
Meanwhile, another 35\% of the excreted P ends up in surface waters or the terrestrial environment through WWTP discharge, diffuses losses from individual autonomous systems, and sewers leaks.
The remaining 15\% is contained in sludge that is incinerated or sent to landfills.
Moreover, while WWTP removal efficiency increased in the 2000s, reaching 80\% on average nationally, it has been followed by a decade of stagnation in every French basin.
The final removal efficiency for each basin (65\% to 85\%) closely correlates with how much of the basin area is defined as P-sensitive in the European directive.
Our results suggest that recycling all P in excretions could help supply 7 to 34\% of the P demand in French food supply, without changing the current food system.
Reshaping agricultural systems (shifting to more plant-based diets, decreasing P losses and food waste) would enable to go even further on the road to food sufficiency.
\end{abstract}

\section*{Keywords}

Wastewater Treatment Plant, Removal efficiency, Wastewater, Sludge, Excretions, Phosphorus

\section*{Abbreviations}

WWTP: WasteWater Treatment Plant $\cdot$ P: Phosphorus $\cdot$  PUE: Phosphorus Use Efficiency $\cdot$ IAS: Individual Autonomous System $\cdot$ N: Nitrogen

\pagebreak

\tableofcontents

\pagebreak

\section{Introduction} \label{sec:introduction}

Phosphorus (P) is a key nutrient for agriculture. Historically, its limited availability has been a major challenge for crop yields, and it is still limiting crop production in a large part of the African continent. At the same time, the excessive phosphorus application in industrial agriculture and the large amounts that flow into urban wastewater treatment plants lead to discharges in rivers and coastal systems. There, the increased phosphate concentration causes eutrophication and can result in harmful algal blooms \citep{Elser2020Phosphorus:, Scholz2014Sustainable}.

Globally, only 20-30\% of P used as fertilizer finally reaches consumers’ plate \citep{Cordell2009story}.
However, if the P stored in agricultural soils is taken into account, the full chain P use efficiency rises up to $\sim$75\% \citep{Sutton2013Our}, leading to a soil P legacy that could be remobilized later along the land–freshwater continuum, worsening eutrophication issues \citep{Sharpley2013Phosphorus}; on the other hand, this remobilized P could also enhance agroecological transitions before full nutrient circularity is achieved, allowing crops to draw from this stock during several decades and decrease phosphate fertilizer applications \citep{LeNoë2020phosphorus}.

These losses mean that P leaving agroecosystems must be replenished by other sources of phosphorus. Phosphate-rock-based fertilizers are a finite resource \citep{White2010Peak, Scholz2013Approaching} and P losses induce eutrophication and biodiversity loss. Moreover, mined phosphate fertilizers contain high concentrations of heavy metals and radioactive elements, which can pollute agricultural soils \citep{Watson2014Mining}.

Even without physical P scarcity, countries consuming P fertilizers can be affected by economically or politically induced P scarcity, following price shocks or geopolitical tensions because of the very skewed distribution of P reserves and production towards a handful of countries \citep{Childers2011Sustainability}. A solution to reduce vulnerability is to close this “broken biogeochemical cycle” \citep{Elser2011broken}, which will also address the environmental issue.

As recovering P is extremely difficult once it has reached the ocean, a sustainable P management requires an almost circular system to bring P emissions into water bodies and landfilling as close to zero as possible.
Besides diffuse P emissions, such as erosion from the soil and especially from agricultural fields during rain events, other main sources of P emissions in Western societies are point sources: human excretions in urban wastewater that reach the environment through Waste Water Treatment Plants (WWTPs). WWTPs generally operate in three steps to address this pollution. Primary treatment removes suspended solids and organic matter, secondary treatment further removes biodegradable matter, and tertiary treatment focuses on P and N enhanced removal \citep{Tchobanoglous2003Wastewater}. Energy recovery from urban wastewater is often emphasized, namely to produce biogas. While this energy can be valuable to operate WWTPs, the potential represents at the very best only 1\% of global anthropic energy consumption \citep{Rittmann2013energy}. More interesting is to recover nutrients for food self-sufficiency: while livestock excretions are almost completely collected and reused in agroecosystems, this is not the case for human excretions.

Studies have consistently found that recycling all P in human excretions could meet $\sim$20\% of the current global P demand in agriculture \citep{Mihelcic2011Global}, well above the estimated 1\% for global energy consumption. Estimations of excretions flows have been produced as part of global P budgets, but with diverging recycling rates, from 10\% to 60\% \citep{Cordell2009story, Liu2008Global, Scholz2014Sustainable, smil2000phosphorus}, due to the uncertainty surrounding the fate of P excreted by the majority of the world population not connected to sewers. 

The uncertainty is much less strong in Western societies, where most people are connected to sewers.
Multiple P budgets of sanitation systems have been produced as part of national P budgets \citep{Häußermann2021National, Hutchings2014nitrogen, Papangelou2021Assessing, Senthilkumar2014Phosphorus}.
Yet they often rely on an average WWTP removal efficiency at the national level, or on mean P concentrations in sewers.
In addition, the flows for people not connected to sewers are not always detailed, and sewers losses are often neglected.

\citet{Esculier2019biogeochemical} have addressed some of these limitations at the regional scale for the Paris mega city area, by relying on operational data of its few WWTPs, and found a P recycling rate of 40\%. Yet the Paris area is quite atypical, with a high population density and large industrial WWTPs. The goal of our work was to expand this regional work to the national scale for France. The main challenge emerging from this is that contrary to the few WWTPs in the Paris area, there are more than 20,000 WWTPs in France. 

In this paper, we describe in detail the French sanitation system to analyze the fate of P in human excreta, in particular how much is recycled to agriculture, and where losses occur. 
This allows to identify the main levers to further increase recycling in the current system.
We also assess the potential of P in human excreta to contribute to food production.

To our knowledge, our work is the first P mass balance of a national sanitation system based on real operational data gathered from the more than 20,000 WWTPs, over one decade or more. Finally, beyond the P mass balance, two interesting points emerge from the building of our dataset: first, how management of P in the sanitation system has changed as the legislation on zones classified as sensitive to eutrophication evolved; second, the strong decrease of domestic emissions of P in detergents, once again following European regulation.

\section{Materials and Methods} \label{sec:materials_methods}

In this section we describe the datasets used for our P budget (outlined in Table \ref{tab:data_sources}) and the specific parameters incorporated into our modeling (detailed in Table \ref{table_P_paremeters}, see also Figure \ref{fig:fig_P_mass_balance_equations} for the detailed calculation of each flow), including a discussion of uncertainties (see Appendix \ref{uncertainties} for more details). We quantitatively estimate uncertainty for each flow with a Monte-Carlo procedure (see Appendix \ref{app:uncertainty-prop}). 

The core of our work consisted of analyzing the data from the 20,000 French WWTPs, issued by the six French Water Agencies: Artois--Picardie, Rhin--Meuse, Seine--Normandie, Loire--Bretagne, Rhône--Méditerranée, Adour--Garonne (Figure \ref{fig:fig_P_basin_removal_efficiency}a). Each water agency is attached to the watershed of its main rivers. 

A P mass balance was then conducted for each of the six water agency basins in France, and the results were consolidated to achieve a national balance. We also present our method to estimate the potential contribution of P in excretions to domestic food supply.
The budget focused on metropolitan France (including Corsica but excluding overseas territories). 

The results are part of a larger project assessing nutrient flows in the French sanitation system; the original data and code to cleanup and analyze the final data are available via an interactive site\footnote{\href{https://thomas-starck.github.io/n-p-sanitation-flows/}{https://thomas-starck.github.io/n-p-sanitation-flows}} and a Zenodo permanent repository\footnote{\href{https://zenodo.org/doi/10.5281/zenodo.7990171}{https://zenodo.org/doi/10.5281/zenodo.7990171}}.

\subsection{Dataset sources}

In France, the number of WWTPs exceeds 20,000, and roughly 1,500 industrial facilities disclose information about their P discharge into sewers. Our estimation of P flows discharged in sewers by industries, WWTPs' inflows and outflows, as well as the various destinations of WWTP sludge, rely on comprehensive datasets detailed hereafter (see also Table \ref{tab:data_sources}). We also address potential data discrepancies by identifying and correcting apparent outliers, and estimate the uncertainties associated with each flow.

\subsubsection{Data corrections}
The data on sludge production, industry discharges, and WWTPs inflows and outflows were scrutinized for obvious outliers, with the following procedure: at the basin level, the P flows for each year were aggregated and scrutinized to identify unexpected spikes in specific years.
This helped in detecting apparent outliers. When it was possible to identify that the peak was simply due to a misplaced comma for that particular year (factor 10, 100, or 1,000), we corrected the entry; otherwise, the outlier was not considered in the analysis and the entry was discarded from the final dataset. For P flows in and out of WWTPs, we identified seven obvious outliers for the 20,000 French WWTPs, over 15--25 years (depending on the water agencies). For sludge production, we identified 25 outliers for the 20,000 WWTPs over the seven years of reported data. For industry discharge, we found three outliers for the 1,000 facilities over one decade of data. The identified outliers and the corrections we made are detailed in our code and our \href{https://thomas-starck.github.io/n-p-sanitation-flows}{website} (e.g. in \textit{WWTP flows preparation $\rightarrow$ 05-Adour-Garonne $\rightarrow$ Data Cleaning}).

\subsubsection{Phosphorus flows in and out of wastewater treatment plants and removal efficiencies}

France counts over 20,000 WWTPs, with data on the annual average P inflows and outflows for each station collected from the six French water agencies. This corresponds to flows A and B on Figure \ref{fig:fig_P_mass_balance_equations}.
While data availability varies across basins (e.g. 1992--2020 for Artois--Picardie, 2009--2020 for Rhône--Méditerranée), our P mass balance relies on mean flows over 2015--2020. P inflow refers to P entering the WWTP (in ton per year). P outflow refers to P discharged by the WWTP in the water bodies. Phosphorus stored in sludge is assumed to be the difference between P inflow and outflow: $\text{P}_\text{sludge} = \text{P}_\text{in} - \text{P}_\text{out}$. Data was solely available for 2015, 2016, 2018 and 2020 in the Seine--Normandie basin, except for 6 of the biggest WWTPs in the Paris region that are managed by SIAAP (Syndicat Interdépartemental pour l'Assainissement de l'Agglomération Parisienne). These 6 facilities handled half of the basin pollution, and we obtained their data for 2007--2020. Uncertainties on these incoming and outgoing flows, discussed in Appendix \ref{A_P_uncertainties_flows}, are estimated to be $\pm$10\%.

The P removal efficiency of a WWTP is defined as: $\text{Efficiency}_P = 1 - \text{P}_\text{out}/\text{P}_\text{in}$.

\subsubsection{Sludge production and destination}

The French sanitation portal database \citep{PortailAssainissementCollectif2023} provides annual reports on sludge production for each WWTP (flow C on Figure \ref{fig:fig_P_mass_balance_equations}), as well as their respective destinations (e.g. direct spreading, composting, incineration, flows D and E on Figure \ref{fig:fig_P_mass_balance_equations}). \citet{Eurostat2022Sewage} also offers a national-scale summary of sludge production.
The total annual production of sludge $p_t$ was between 1 and 1.1 Mt of dry matter, consistent over the years. From 2018 to 2021, where sludge destinations are stable and consistent, approximately 45\% of sludge was composted and 30\% directly spread on crops. Uncertainties, discussed in Appendix \ref{A_P_uncertainty_sludge}, are estimated to be $\pm$10\%.

\subsubsection{Large industries discharge in sewers}

Some industries release their P pollutant in the sewers, after a preliminary treatment (flow H in Figure \ref{fig:fig_P_mass_balance_equations}). To estimate the annual discharge of P by industries into sewers, we used the GEREP database, provided by the Direction Générale de la Prévention des Risques from the Ministry of Ecological Transition, which reports approximately 1,200 facilities. Uncertainties, discussed in Appendix \ref{A_P_industries}, are estimated to be $\pm$10\%.

\subsubsection{French population}

We used data from the French National Institute of Statistics and Economic Studies \citep{INSEE2022Population}, which provides information on the population in 2018 in each city together with their age distribution. The uncertainty associated with population data is probably negligible in comparison to the other parameters given the large spatial scale. However, there is a $\pm$3\% uncertainty at the national level as we do not account for tourism and demographic changes; this is discussed in Appendix \ref{A_P_population}.

\subsubsection{Reference years for the P budget}

We calculated the P balance flows for each water agency, using average quantities of industry discharges to sewers and WWTP P flows over the 2015--2020 period. To determine sludge destination, we took an average of the years 2018--2021, as data before this period is inconsistent and barely reported.

\subsection{Individual parameters}

This section outlines the various coefficients that were used to calculate the phosphorus mass balance (Table \ref{table_P_paremeters}, Figure \ref{fig:fig_P_mass_balance_equations}), along with their corresponding uncertainties (Table \ref{tab:uncertainties_parameters}).

\begin{table}[h!]
    \caption{Parameters used for the P mass balance of the sanitation system. Refer to Figure \ref{fig:fig_P_mass_balance_equations} to see the computation for each flow of Figure 1. IAS: Individual Autonomous System.}
    \label{table_P_paremeters}
    \centering
    \begin{tabular}{m{0.35\linewidth}m{0.27\linewidth}M{0.12\linewidth}M{0.16\linewidth}}
        \toprule
            \multicolumn{2}{c}{\textbf{parameter}} & \textbf{value} & \textbf{source} \\ 
        \midrule
            \multirow{8}{=}{Direct discharge and losses from sewers at the basin scale ($\%$ of pollution entering the wastewater treatment plants)} & Artois--Picardie basin & $20\%$ & \multirow{7}{=}{\centering Water agencies data} \\
             & Rhin--Meuse basin & $20\%$ & \\
             & Seine--Normandie basin & $10\%$ & \\
             & Loire--Bretagne basin & $15\%$ & \\
             & Adour--Garonne basin & $7\%$ & \\
             & Rhône--Méditerranée basin & $7\%$ & \\
             & France & $10\%$ & Combined basins \\
        \hline
            \multirow{8}{=}{Total population (million) and share not connected to sewers} & Artois--Picardie basin & 4.8M, $15\%$ & \multirow{7}{=}{\centering Water agencies data } \\
             & Rhin--Meuse basin & 4.3M, $6\%$ & \\
             & Seine--Normandie basin & 19M, $7\%$ & \\
             & Loire--Bretagne basin & 13M, $24\%$ & \\
             & Adour--Garonne basin & 7.8M, $30\%$ & \\
             & Rhône--Méditerranée basin & 16M, $25\%$ & \\
             & France & 64.9M, $15\%$ & Combined basins \\
        \hline
            \multirow{3}{=}{IAS P mass balance} & \multirow{1}{=}{sludge} &  \multirow{1}{=}{\centering$15\%$} & \cite{Catel2017Inventaires}\\
            & \multirow{1}{=}{underground diffuse losses} & \multirow{1}{=}{\centering$85\%$} & \cite{Risch2021Applying}\\
        \hline
           \shortstack[l]{For people not connected to sewers, share of excretions in sewers} & & \multirow{2}{*}{20\%} & \multirow{2}{*}{estimation} \\ 
           \shortstack[l]{(i.e. excretions outside home, in public spaces connected to sewers).}\\
        \hline
            \multirow{10}{=}{P ingestion by French people (gP/day)} & female, $\leq$10 years old & 0.96 & \multirow{10}{=}{\centering INCA3 study \citep{dubuisson2019third, data.gouv2021Données}} \\
             & female, 11-18 years old & 1.09 & \\
             & female, 18-44 years old & 1.11 & \\
             & female, 45-64 years old & 1.12 & \\
             & female, $\geq$65 years old & 1.03 & \\
             & male, $\leq$10 years old & 1.01 & \\
             & male, 11-18 years old & 1.32 & \\
             & male, 18-44 years old & 1.41 & \\
             & male, 45-64 years old & 1.42 & \\
             & male, $\geq$65 years old & 1.32 & \\
        \bottomrule
    \end{tabular}
\end{table}

\subsubsection{Phosphorus excretions (urine and feces)}

We used data from the INCA3 study \citep{dubuisson2019third, data.gouv2021Données} to determine P ingestion by French citizens, disaggregated by age and sex categories. This is coupled with INSEE data describing the French population by city and by age \citep{INSEE2022Population}, to determine P excretions for each water agency basin and France (flow I in Figure \ref{fig:fig_P_mass_balance_equations}). This resulted in an average national P excretion of 0.44 kgP$\cdot$cap$^{-1}\cdot$year$^{-1}$. More information and a discussion of the plausible range (0.44--0.58 kgP$\cdot$cap$^{-1}\cdot$year$^{-1}$) based on our literature review, and used as an uncertainty range, can be found in Appendix \ref{A_P_uncertainty_excretions}.

\subsubsection{Individual Autonomous Systems}\label{sect_IAS}

To estimate the share of people with individual autonomous systems (IAS) not connected to sewers, we used the figures from the Status Reports (“Etat des lieux” in French) of 6 water agency basins. The resulting weighted mean was 15\%. This is in line with the Ministry of Ecological Transition, which stated that 15 to 20\% of the population is not connected to sewers and relies on Independent Wastewater Treatment \citep{ministere2023}
In addition, we assumed that individuals not connected to sewers excrete 20\% of their P in public spaces connected to sewers \citep{Hellstrand2015Nitrogen}. This allowed to compute flows J and K in Figure \ref{fig:fig_P_mass_balance_equations}.

\citet{Catel2017Inventaires} and \citet{Risch2021Applying} produced a P mass balance of septic tanks. Based on their results, we assumed that for people with individual autonomous system, 85\% (80-90\%) of the P ends as underground diffuse losses (effluent of the septic tanks), and 15\% (10-20\%) in septic tank sludge, which is then brought to WWTPs for further treatment (flows O and N in Figure \ref{fig:fig_P_mass_balance_equations}). The uncertainties are discussed in Appendix \ref{A_P_uncertainty_individual_system}.

\subsubsection{Direct discharge losses}

A portion of the discharged phosphorus doesn't reach the WWTPs because of incorrect pipe connections, sewer overflows and leaks. Through individual WWTP data and water agency status reports we established a national weighted average of 10\% for pre-WWTP loss (flow G in Figure \ref{fig:fig_P_mass_balance_equations}). There is a large factor 2 uncertainty of this parameter, with plausible values between 5 and 20\%, discussed in Appendix \ref{A_P_loss_before_WWTP}.

\subsubsection{Calculation of residual pollution}

From the inputs measured at the WWTP inlet and our estimated values of network direct discharges, we obtained the theoretical flow entering sewers. This value is greater than the combined discharge of industry and human excreta into sewers. We call the difference ``residuals'' (flow L on Figure \ref{fig:fig_P_mass_balance_equations}), which includes several elements: household detergents, household kitchen leftovers, small industries not required to declare discharges into sewers and surface runoff entering the combined sewer networks.

To estimate the residual flow entering individual autonomous systems (flow M on Figure \ref{fig:fig_P_mass_balance_equations}), we made the assumption that the residual flow of the networks is essentially due to domestic discharges, and applied the same residues-to-excretions ratio to individual autonomous systems (still taking into account that people with individual autonomous system excrete 20\% of their P in public spaces connected to sewers, see section \ref{sect_IAS}).

\subsection{Potential for domestic food consumption}

We assessed the potential contribution to food security if all P in excretions were reused as fertilizers. First, we compared the amount of P in excretions to P fertilizers consumption reported by \citet{FAOSTAT2023Fertilizers}. This is useful to indicate the potential decrease in dependence towards providers of phosphate rocks. However, this does not indicate how much of the P in domestic food consumption could be covered. Indeed, France is largely an open food system \citep{Billen2018Two}, and some P fertilizers are used to grow crops that are finally exported, while others are hidden in imported crops and not accounted for. 

To address this limitation, we used the notion of P footprint, to get a better sense of the potential contribution of human P excretions to food supply in France. The P footprint allows to see the required P to feed a population, and accounts for the requirements beyond the country's borders that may be hidden by food commodity imports or exports. The P footprint (P$_f$) is associated to the food supply P content (P$_s$) of a population and the P use efficiency (PUE) through the following relation:  $\text{P}_f = \text{P}_s / \text{PUE}$. Note that the food supply refers to food before waste at the consumer stage, so its P content is higher than the P ingested and excreted. 

We then compared the P content of excretions to the footprint of the domestic food supply, to see what share could be covered if all of the P in excretions were reused as fertilizers. We also distinguished between plant- and animal-based products since they have very different PUEs (and footprints) on average. We looked at what would happen if all P excretions used as fertilizer were prioritized towards crops aimed at directly feeding humans. This gives an upper bound for the contribution of excreted P to food production (and thus domestic food self-sufficiency), since crops PUE is much larger than animal products PUE. More details about the calculations can be found in Appendix \ref{P_footprint} and the Supplementary Spreadsheet.

We computed these contributions for the Paris megacity --- 10 million inhabitants, taken as a proxy for France, from data in \citet{Esculier2019biogeochemical}. Though the precise results at the national scale might be slightly different than for Paris (e.g. P intake of 0.44 kgP$\cdot$cap$^{-1}\cdot$year$^{-1}$ in our study for France vs 0.43 in \citet{Esculier2019biogeochemical} for Paris megacity), we make the assumption that they should be close since the diet should be similar throughout the country. 

We also looked at figures for the whole world, based on \citet{Sutton2013Our}, for two main reasons. First, it allowed us to have a sanity check on our P footprint method, and to compare it to results already present in the literature. Second, beyond the current food system, the potential contribution of excretions to food security could be further increased by redesigning the food system (reduced share of animal products, increased PUE, decreased food waste). Comparing France to the world is a way to evaluate the effect of diets without further detailed modeling. 

All values used as inputs for our analysis came from the values of \citet{Sutton2013Our} and \citet{Esculier2019biogeochemical}; they are reported in Appendix \ref{P_footprint} and the Supplementary Spreadsheet.

\section{Results}

\subsection{French sanitation system P budget and recycling to agroecosystems}

\begin{figure}[h!]
\centering
    \includegraphics[width=0.95\linewidth, trim = {0 0 0 0}, clip]{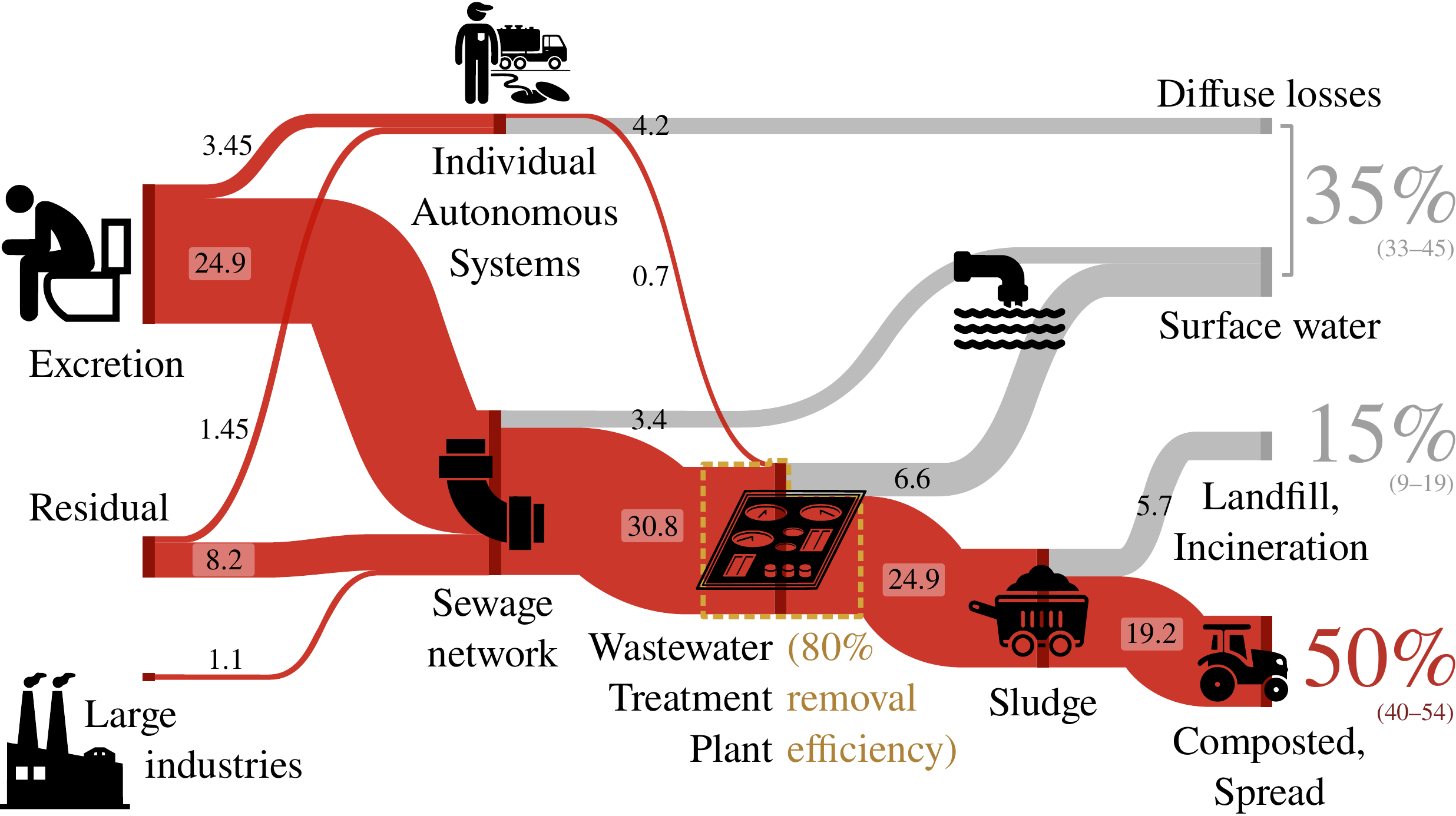}
    \caption{Yearly phosphorus flows in the French sanitation system for the years 2015 to 2020 (ktP).
    The average removal efficiency of the WWTP is around 80\%, leading to approximately 35\% of phosphorus loss in surface water and diffuse losses (95\% confidence interval: 33--45\%), 15\% loss to landfill or incineration (CI: 9--19\%) and 50\% returned to agricultural land (CI: 40-54\%).
    Detailed uncertainties for all flows are presented in Appendix \ref{uncertainties}.}
	\label{fig:fig_P_sankey}
\end{figure}

For the period 2015--2020, we estimated that French P excretions totaled 28.4 kt P$\cdot$year$^{-1}$, with large industries discharging 1.1 kt P$\cdot$year$^{-1}$ into sewers and residual P (e.g. detergents, kitchen waste, runoff, small industries) amounting to 9.6 kt P$\cdot$year$^{-1}$ (Figure \ref{fig:fig_P_sankey}). The 39.1 kt P resulting from these three flows enter the sanitation system through sewers and individual autonomous systems. Of this, 19.2 kt P ($\sim$50\%) is used as fertilizer for crops (30\% as direct application of WWTP sludge and 20\% after composting), 14.2 kt P ($\sim$35\%) reaches surface water or is diffusely lost in the ground, and 4.2 kt P ($\sim$15\%) is incinerated or lost in landfills.

P discharged by WWTPs to surface waters is 6.6 ktP (6.0--7.2 kt P). The combined losses from individual autonomous systems -- 4.2 kt P (3.5--6.8 kt P) -- and direct discharge from sewers -- 3.4 kt P (1.8--7.6 kt P) -- add up to 7.6 ktP (2.5--8.7 kt P), thus exceeding WWTP discharge, but with much higher uncertainty. Because of these losses, despite a national WWTP removal efficiency of 80\%, and 75\% of French sludge being used as crop fertilizer (combined recovery of $80\% \cdot 75\% = 60\%$), the whole sanitation system recycling rate is only 50\%.

\subsection{WWTP removal efficiency through space and time}

\begin{figure}[h!]
\centering
    \includegraphics[width=0.95\linewidth, trim = {0 0 0 0}, clip]{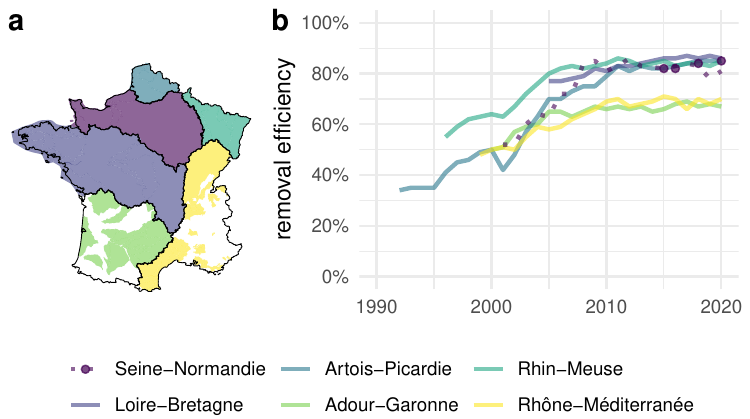}
    \caption{a) Phosphorus sensitive areas of the six French water agencies basins, as of 2024. Colored areas are considered P-sensitive, white areas are considered non-sensitive. b) Temporal evolution of P removal efficiency for the six basins. For Seine--Normandie, complete data was available only in 2015, 2016, 2018 and 2020 (points). The dotted line represents the partial data from the six largest Seine--Normandie facilities, handling about half of the total pollution.}
	\label{fig:fig_P_basin_removal_efficiency}
\end{figure}

Efficiency of P removal in each French water basin increased during the 2000s but reached a plateau in the 2010s at a national mean of 80\%. The final performance correlates closely to the P sensitive area classification; Adour--Garonne and Rhône--Méditerranée basins, partly considered ``non-sensitive to P'' (Figure \ref{fig:fig_P_basin_removal_efficiency}a), have lower removal efficiencies of 65--70\% (Figure \ref{fig:fig_P_basin_removal_efficiency}b), while the other basins achieve $\sim$80--85\% efficiency.

\begin{figure}[h!]
\centering
    \includegraphics[width=0.95\linewidth, trim = {0 0 0 0}, clip]{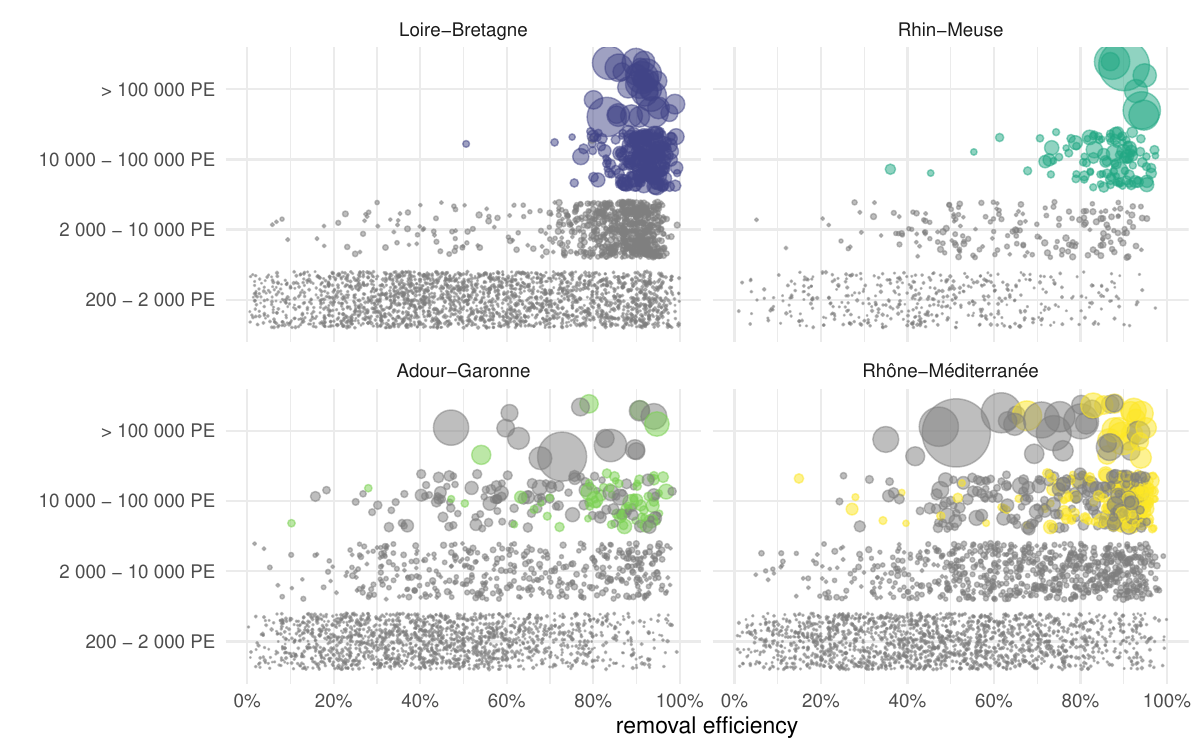}
    \caption{Individual WWTP P removal efficiencies, averaged over 2015-2020, in four of the French water agencies basins. Each dot is a WWTP, size is proportional to nominal capacity, expressed in population equivalent (PE). Dots are highlighted if the WWTP must reach minimal performances due to P sensitive area classification (see Figure \ref{fig:fig_P_basin_removal_efficiency}a and text).}
	\label{fig:fig_P_WWWTP_removal_efficiency}
\end{figure}

The primary reason for this contrast is the removal efficiency of large WWTPs. French legal enactments \citep{Légifrance2015Arrêté} requires facilities larger than 10,000 in areas classified as sensitive to P to have 80\% annual P removal or outflow concentration below 1--2 mg P$\cdot$l$^{-1}$. These make up 5\% of French WWTPs but handle 80\% of flows and are responsible for most of the differences between basins. There is no automatic requirement for smaller WWTPs or non-P sensitive areas. Figure \ref{fig:fig_P_WWWTP_removal_efficiency} illustrates the effect of the legislation. Large WWTPs in P-sensitive areas tend to have removal efficiencies $>$80\%. However, in Adour--Garonne and Rhône--Méditerranée basins, large facilities not classified as P sensitive frequently have lower removal efficiencies. The removal efficiencies of small ($<$2,000 population equivalent) WWTPs take a wide range of values between 0 and 100\%. Interestingly, intermediate WWTPs (2,000--10,000 population equivalent) tend to have higher removal efficiencies than small ones, since they are sometimes subject to additional local legal requirements.

\subsection{Current agricultural potential}

For 2015--2020, P fertilizer consumption in France was 180--200 kt P$\cdot$year$^{-1}$, so the 28.4 kt P$\cdot$year$^{-1}$ excreted could replace $\sim$15\% of the fertilizer consumption. Consolidated global figures for 2000--2010 by \citet{Sutton2013Our} are 14--18 MtP for fertilizer consumption and 4 MtP in excretions, resulting in a potential of $\sim$25\%. This is an indication of how much independence could be gained from phosphate rocks imports.

\begin{figure}[h!]
    \centering
    \includegraphics[width=0.95\linewidth, trim = {0 0 0 0}, clip]{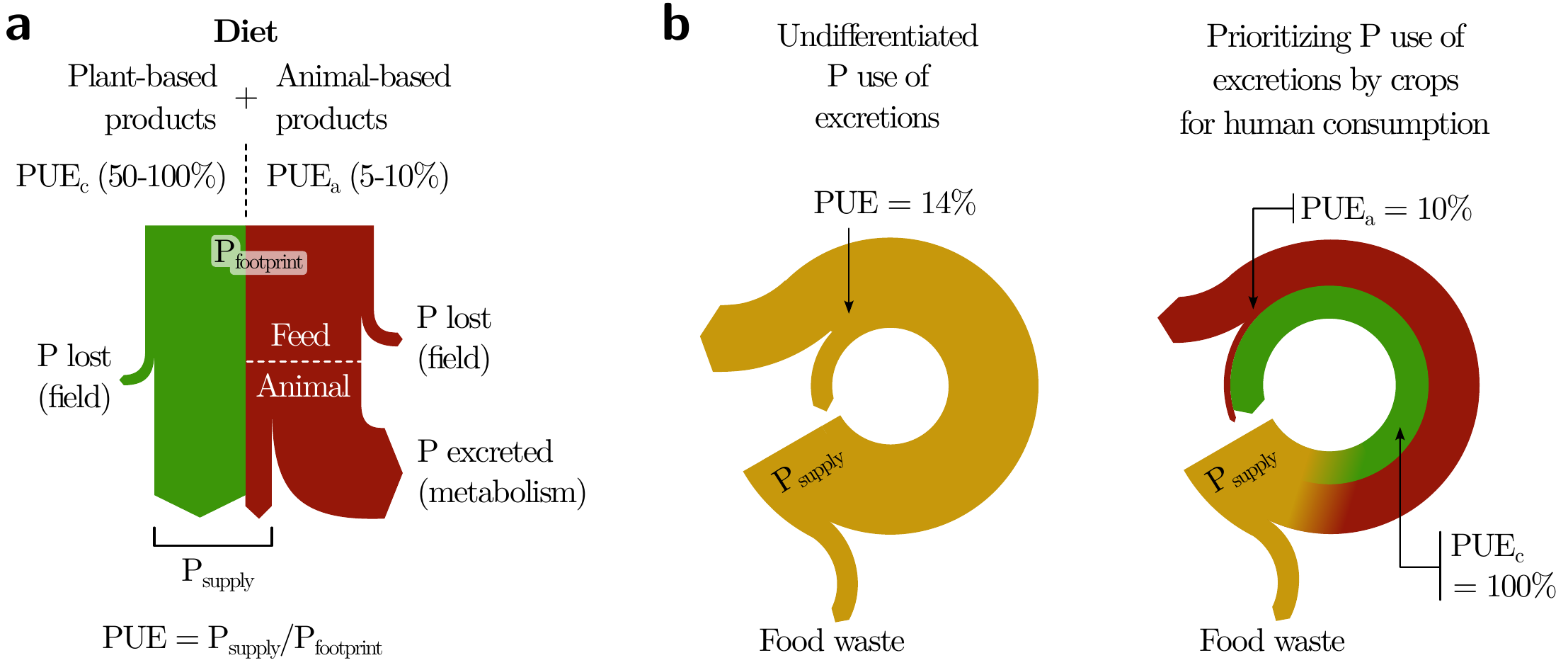}
    \caption{\textbf{a}. Phosphorus Use Efficiency (PUE) of plant-based (crops) and animal products for human consumption. The PUE of plant-based products is much higher than that of animal-based products which incurs both losses from crops in the field and even more significant losses due to the animal metabolism. The PUE of a diet distributed between a fraction $f_c$ of plant-based products and $f_a = 1 - f_c$ of animal-based products is given by $\text{PUE} = 1 / (f_c / \text{PUE}_c + f_a / \text{PUE}_a)$. \textbf{b}. The contribution of P in excretions to food supply can be improved by prioritizing the fertilization of crops for human consumption. For instance, with a diet composed of 30\% plant and 70\% animal products, a crop PUE$_c$ of 100\% and an animal-products PUE$_a$ of 10\%, the resulting average PUE is 14\%. The use of excreta to grow crops for human consumption thus yields more P in food supply than an indiscriminate use including animal feed. Note that in reality ingestion of sea products also contribute to N excretions.}
    \label{fig:pue}
\end{figure}

At the global scale, based on our P footprint method, we found that food production requires approximately 21--36 MtP, meaning that completely recycling human P excretions could cover around 10--16\% of the P needed for food supply in the current food system. Prioritizing the recycling to plant-based production for human consumption could increase this rate to 40\%. In the case of France (extrapolated from the data of the Paris megacity), where the fraction of animal-based products in the diet is larger than the global average, the potential contribution to food domestic supply is only about 7\% but it could go up to 34\% if  plant-based products for human consumption are prioritized (Table \ref{table_P_footprint} and Figure \ref{fig:pue}b).

\section{Discussion} \label{Discussion}

\subsection{Consistency of P removal efficiencies with the literature}

The 80\% removal efficiency at the basin level in P sensitive areas during 2010--2020 are in line with \citet{Degrémont2005Mémento}, who reported 80--90\% removal efficiency for P tertiary treatment in WWTPs. \citet{Vigiak2020Domestic} and \citet{Drecht2009Global} propose respectively 30\%-60\%-90\% and 10\%-45\%-90\% removal efficiencies for primary, secondary, and tertiary treatments. Given that 1--2\% of the French urban wastewater is handled with primary treatment, 15--20\% with secondary, and 80--85\% with tertiary \citep{Eurostat2023Population}, the national removal efficiency would be 85\% for \citet{Vigiak2020Domestic} and 82\% for \citet{Drecht2009Global}, which is close to our study's 80\%.

Regarding the temporal evolution of removal efficiency, \citet{Drecht2009Global} suggested a P removal efficiency of 44\% in 1990 and 59\% in 2000 for Europe. For basins with data at those times, our values are $\sim$40\% in 1990 and 50--63\% in 2000.

\subsection{Regulations led to increased P removal efficiency but not to higher recycling}

P removal efficiencies closely match classification into P sensitive areas and European directives requirements. First, with the temporal increase in the 2000s for each basin and the stagnation levels in the 2010s (Figure \ref{fig:fig_P_basin_removal_efficiency}); second with the effect on large WWTPs when they are in sensitive areas (Figure \ref{fig:fig_P_WWWTP_removal_efficiency}). Further regulation might decrease P discharge from WWTPs in Southern basins and for smaller facilities. However, about as much loss happens outside WWTPs, because of sewer losses and individual autonomous systems. Improvements in WWTP removal efficiency will only increase sewers and individual autonomous systems losses relative share. Thus, future regulation on P pollution may need to look beyond WWTPs and address the whole sanitation system.

\subsection{France recycling rate is one of the highest among western countries}

Besides removal efficiencies, France recycling of 50\% of P in human excreta does not appear very high but is higher than other Western countries, whose recycling rates around 2010-2020 are estimated to be (non-exhaustive list): below 15\% for Belgium \citep{Papangelou2021Assessing}, ~25\% for Austria \citep{Egle2014Austrian}, ~50\% for the UK \citep{Cooper2013substance}, ~25\% for Sweden \citep{Linderholm2012Phosphorus}, ~25\% for Germany \citep{Jedelhauser2015Losses}, virtually null for Netherlands \citep{Smit2015substance}, and 35\% in Japan \citep{Matsubae-Yokoyama2009Material}. All these countries have relatively high removal efficiencies (from 60\% to 90\%). So in these Western and industrialized countries, the main factor underlying recycling is not the removal efficiency  but the share of WWTP sludge reused as fertilizer.

\subsection{No silver bullet to increase recycling rate in the current sanitation system, but several levers}

In the case of France, currently 75\% of WWTPs sludge is spread on agricultural land. Recycling all P in produced sludge could increase the sanitation recycling rate of P by 15 percentage points, or 30\% in relative terms (Figure \ref{fig:fig_P_sankey}). Increasing WWTP removal efficiencies is another lever, especially in the Southern basins. At best, hypothesizing 100\% removal efficiency, it would increase recycling by 6.6 kt P$\cdot$year$^{-1}$ (6.0--7.2 kt P$\cdot$year$^{-1}$), or about 15--20 percentage points. The potential improvements are similar to the P lost from sewers (leaks and storm water overflows) and individual autonomous system together, but these flows are overlooked in current policies. Finding ways to recover P from individual autonomous systems and reducing sewer losses should be considered by policy makers if the aim is to further improve P circularity.

However, activating all these levers would still leave some issues. First, this will not enhance N cycling in the sanitation system, which, contrary to P, is dissipated in the air as $\text{N}_\text{2}$ and $\text{N}_\text{2}\text{O}$ when treated in WWTPs. Second, future regulation and lower thresholds for contaminants (especially heavy metals) in WWTP sludge may prevent higher recycling rates. Source separation of excreta, before they reach sewers, could be one way to address these issues \citep{larsen2013source}.

\subsection{P in excretions could currently cover 7 to 34\% of French food supply}

\begin{table}[h!]
    \caption{Phosphorus footprint from plant and animal products implies different potential contribution of P excretions to P food supply depending on what is fertilized. In the current agricultural system, prioritizing crops for human consumption enables larger contributions to the P food supply, both in France and globally. See Table \ref{suptable_P_footprint} for detailed figures sources}
    \label{table_P_footprint}
    \centering
        \begin{tabular}{m{0.3\linewidth}>{\centering\arraybackslash}p{0.3\linewidth}>{\centering\arraybackslash}p{0.3\linewidth}}
        \toprule
                 & \shortstack{\textbf{World (MtP)}\\(years 2000--2010)} & \shortstack{\textbf{Paris Megacity (ktP)}\\(year 2012)} \\ 
            \hline
                P in food supply & 5.1 & 6.6\\
 P in ingestion/excretions& 4&4.3\\
 \hline
 crops PUE& 50\%&105\% (mining)\\
 animal products PUE& 5-10\%&7\%\\
 total PUE& 12-21\%&11\%\\\hline
                \multirow{5}{=}{P in food supply (before food waste) obtained with P excretions used as fertilizer}& \textit{Prioritizing plant production} & \textit{Prioritizing plant production} \\
                 & 2& 2.3\\
                 & & \\
                 & \textit{Not prioritizing plant production} & \textit{Not prioritizing plant production} \\
                 & 0.5--0.8& 0.5\\
            \hline
                \multirow{5}{=}{Share of P consumption covered by using excretions as fertilizer}&\textit{Prioritizing plant production} & \textit{Prioritizing plant production} \\
                 & $40\%$ (including fish)& $34\%$ (including fish)\\
                 & & \\
                 & \textit{Not prioritizing plant production} & \textit{Not prioritizing plant production} \\
                 & 10--16\% (including fish)& $7\%$ (including fish)\\
            \bottomrule
        \end{tabular}
\end{table}

The current global recycling of P excretion is unclear, with estimations from 10\% to 66\% \citep{Cordell2009story, Liu2008Global, Scholz2014Sustainable, smil2000phosphorus}, owing to uncertainties concerning how human waste is handled for people not connected to sewers, representing the majority of world population \citep{Drecht2009Global}. However, it has consistently been estimated that recycling all human P excretions could cover roughly 20\% of current P fertilizer demand in the current food system \citep{Mihelcic2011Global}. 

The P footprint method applied to the global food system gives a P footprint distribution of 16--28\% for plant products and 72--84\% for animal products (see Table S3, Section S4 and supplementary spreadsheet for computation details). This is in line with the assertion of \citet{Sutton2013Our} that ``globally, the 80\% of N and P in crop and grass harvests that feeds livestock ends up providing only around 20\% (15--35\%) of the N and P in human diets''. Our global food PUE (plant + animal, before waste) is 12--21\% (Table \ref{table_P_footprint}), consistent with the reported 12--20\% full-chain PUE \citep{Sutton2013Our}. This consistency of the P footprint method at the global level thus comforts our results at the French scale.

Importantly, this P footprint method enables to go beyond the potential for food self-sufficiency in the current food system (10--16\% globally and 7\% in France): it suggests that prioritizing the recycling towards plant production for human consumption could help cover up to 40\% of the P food supply globally and 34\% in France. These results are relevant from a food security or a geopolitical perspective, to reduce national dependence on foreign phosphate rock supply.

It is worth mentioning that these figures concern the food system as currently designed. The rates could also be further increased by reducing food loss and waste, increasing crop and animal PUE and shifting towards more plant-based diets. While increasing crop PUE (currently around 50\%) at the global scale is possible, this cannot be considered for France which is already in a situation of P mining from soils. Interestingly, even though France has a crop and an animal PUE twice as good as the global figure (100\% vs 50\% and 10\% vs 7\%), the total food PUE is higher at the global scale than in France (respectively 12--21\% and 11\%). This is because the ratio of P sourced from plant-based versus animal-based products is 2 at the global scale but 0.43 in France, highlighting the impact of mostly animal-based diets in Western countries. About a third of the global food production is lost or wasted \citep{Gustavsson2011Global}, and reducing it would further increase the full chain PUE and the contribution of P excretions to food supply. 

\subsection{Regulations on P in detergents contributed to the decrease of P discharge by reducing P inputs to WWTPs}

Figure \ref{fig:fig_P_basin_removal_efficiency}b shows the improvements in WWTPs P removal efficiencies, which was an important factor in reducing P discharge to water streams in the last two decades. Another factor of similar importance was the progressive ban on phosphate in detergents (laundry and dishwater) which, in the 90s, was a flow of similar magnitude as P in excretions.

At least three successive limitations on P in detergents were taken in France and in Europe. In France, phosphorus in laundry detergents were banned in 2007 \citep{Légifrance2007Décret}. In 2013 the European Union banned consumer laundry detergents with standard doses $>$0.5 gP, and in 2017 automatic dishwater detergents with standard doses $>$0.3 gP \citep{EU2012Regulation}. As roughly 75\% of the P in detergents came from laundry \citep{Drecht2009Global}, most of the effect of the EU ban is thus expected to be seen before 2013. Supposing a consumption of 1 or 2 ``standard doses'' per capita and per week for both dishwater and laundry would give an upper limit P detergent emission of 0.04--0.08 kgP$\cdot$cap$^{-1}\cdot$year$^{-1}$ after 2017, or 7--13\% of our computed national emissions (0.6 kgP$\cdot$cap$^{-1}\cdot$year$^{-1}$).

For the 2000s in EU25, \citet{Drecht2009Global} proposed a human P emission in wastewaters (domestic + industries) of 0.9--1 kgP$\cdot$cap$^{-1}\cdot$year$^{-1}$, 40\% of which were due to detergents from laundry and dishwashers (0.4 kgP$\cdot$cap$^{-1}\cdot$year$^{-1}$). Despite large variations between European countries, this is in line with \citet{Vigiak2020Domestic}, based on \citet{Bouraoui2009Nutrient}, reporting a detergent consumption of 0.37 kgP$\cdot$cap$^{-1}\cdot$year$^{-1}$ in 2005 for France. These values imply a non-detergent human P emission (domestic + industries) of $\sim$0.6 kgP$\cdot$cap$^{-1}\cdot$year$^{-1}$. This corresponds to our figure of 0.6 kgP$\cdot$cap$^{-1}\cdot$year$^{-1}$ for 2015--2020 in France, after the ban on P detergents.  

Our estimation of residual P discharge is 9.6 ktP, or 0.15 kgP$\cdot$cap$^{-1}\cdot$year$^{-1}$ (Figure \ref{fig:fig_P_sankey}), but this comes with high uncertainties, combined from those on sewer losses and human excretions. Following our Monte-Carlo procedure, the 95\% uncertainty range associated with residual emission goes from 0 to 15.2 ktP (0--0.23 kgP$\cdot$cap$^{-1}\cdot$year$^{-1}$).  

From our data, it is not possible to directly observe the impact of P detergent ban through the temporal evolution of P emissions per capita. Reporting rates of WWTP data increase through time, reaching full levels in the 2010s, but are still incomplete before. This complicates the estimation of the temporal evolution of the number of people connected to sewers associated with the reported incoming pollution. Data about sewers overflows and leakage are also non-existent before 2015.

\begin{figure}[h!]
\centering
    \includegraphics[width=0.95\linewidth, trim = {0 0 0 0}, clip]{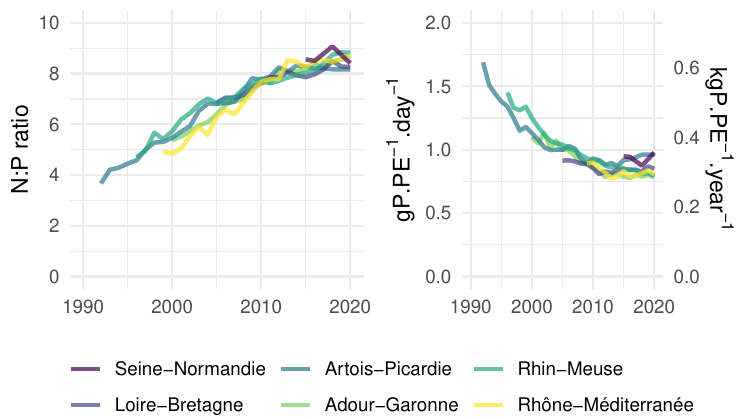}
    \caption{For each of the French 6 water agencies basins: a) N:P molar ratio of incoming flows in WWTPs and b) incoming P in WWTPs, normalized by WWTPs capacity, in population equivalent (PE). In 2020, the total French capacity was 100 million PE, for 52.5 million inhabitants connected to sewers.}
	\label{fig:fig_P_ratios}
\end{figure}

However, we can normalize incoming P flows by the nominal capacity of the reported WWTP data (in 2020 France total nominal capacity was 100 million population equivalents, for 54 million people effectively connected to sewers). We also normalize by incoming N flows (Figure \ref{fig:fig_P_ratios}). In both cases, most of the decrease in incoming P loads happened before 2010, and by 2015 the stabilization was over, consistent with the EU directive. These two ratios could be used in further work as inputs to model emissions of P detergents for 1990--2020 in France. Note that the trends are consistent between the independent datasets of different basins, which suggests that the results are reliable.

\subsection{Limitations}

The main limitations of the study are uncertainty in individual autonomous systems' P mass balance and rough estimation of sewer losses. Uncertainty in individual autonomous systems will be difficult to reduce, unless a large part of these facilities (5 million in France) is regularly monitored, which is implausible. The issue is different for sewers direct discharge, some of which are monitored. Increasing reporting rates may help reduce our current high uncertainty on this matter ($\sim$50\%). Combining this with quantification of the P flow due to runoff entering combined sewers networks could further improve our P budget estimations. There are also uncertainties concerning P ingestion and excretion, which could be further assessed.

\section{Conclusion} \label{Conclusion}

Though France has a high P reuse in agriculture, with a 50\% recycling rate, this is not primarily due to a will to recycle but due to a will to limit P discharge in surface water, combined with a permissive legislation on sewage sludge. Enhancing WWTP P-removal in basins where surface water is currently considered “non sensitive” to eutrophication could both reduce water pollution and increase P recycling rate.

However, sewage sludge use in agriculture also presents significant drawbacks due to health and environmental impacts, notably from various pollutants and heavy metals.
Because of this, future legislation may restrict its use, thus reducing nutrient cycling. Source separation of human excreta is therefore an important method to investigate, as a way to further increase P recycling while preventing it from mixing with contaminants in the sewers.

Redesigning the food system to reduce food waste and losses, and to increase P use efficiency across the production chain, together with a switch to more plant-based diets could further increase the potential of P excretions used as fertilizers.

More precise assessments of the losses occurring in sewage networks -- notably through more systematic direct monitoring of sewers -- would be necessary to reduce the uncertainties for their P mass balance.

\section*{Declarations} \label{Declarations}

\subsection*{Funding}

Thomas Starck was awarded a PhD scholarship from Ecole Polytechnique and Ecole des Ponts Paristech. 

Tanguy Fardet was awarded a Postdoctoral Fellowship from the Marie Skłodowska-Curie Action of the European Commission. 

The \href{https://www.leesu.fr/ocapi}{OCAPI Program} is funded by several public institutions. The funders had no role in study design, data collection and analysis, decision to publish, or preparation of the manuscript.

\subsection*{Competing interests} None

\subsection*{Ethics approval} Not applicable

\subsection*{Consent to participate} Not applicable

\subsection*{Consent for publication} Not applicable

\subsection*{Availability of data and code} The results are part of a larger project assessing nutrient flows in the French sanitation system; the code to generate, cleanup, and analyze the data is available as \href{https://codeberg.org/TStarck/N\_P\_France\_sanitation\_system}{an interactive website}. In addition to the code, the original datasets sources before cleanup, the consolidated data used for the analysis, the graphs and more information are available on Zenodo: \href{https://doi.org/10.5281/zenodo.7990171}{https://doi.org/10.5281/zenodo.7990171}.

\subsection*{Authors' contributions}  

Thomas Starck: Data curation, Formal analysis, Methodology, Funding acquisition, Software, Visualization, Writing - Original draft, Writing - review \& editing

Tanguy Fardet: Data curation, Validation, Methodology, Visualization, Funding acquisition, Writing - Original draft, Writing - review \& editing

Fabien Esculier: Supervision,  Project administration, Methodology, Writing - review \& editing

\subsection*{Acknowledgement}
The Sankey diagramm (Figure \ref{fig:fig_P_sankey}) was made using the Open-Sankey tool https://open-sankey.fr/. On the Sankey diagram, the wastewater treatment plant icon is adapted from Daan's work; other icons were made by Luis Prado, Marco Livolsi, Jae Deasigner, Singlar, Gan Khoon Lay, WEBTECHOPS LLP, Trevor Dsouza, and sandra from the Noun Project (\href{https://thenounproject.com}{https://thenounproject.com}).

\pagebreak

\appendix
\counterwithin{figure}{section}
\counterwithin{table}{section}

\section{Appendices}

\subsection{Equations and workflow of the P mass balance}\label{P_mass_balance}

\begin{figure}[h!]
\centering
    \includegraphics[width=0.92\linewidth, trim = {0 0 0 0}, clip]{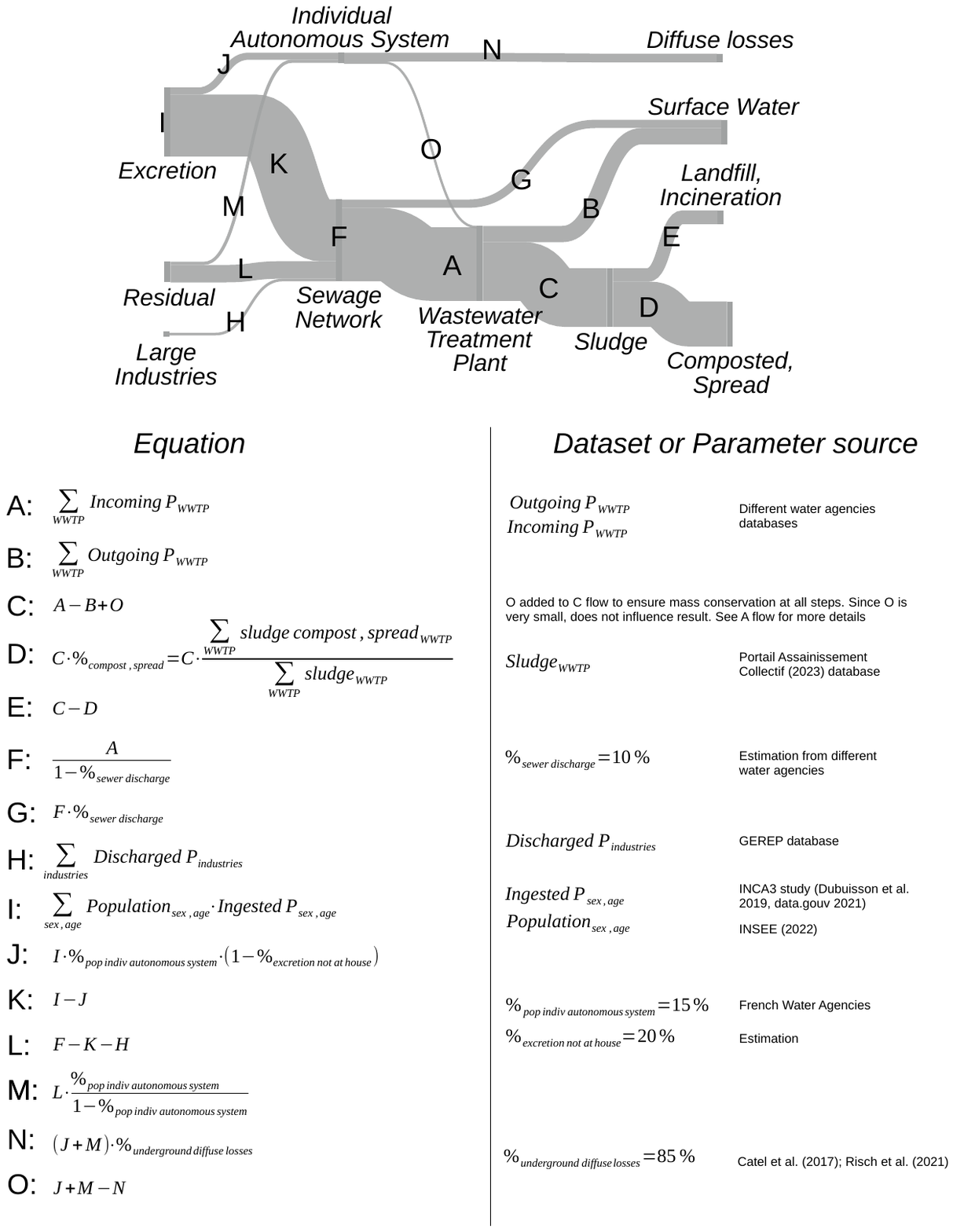}
    \caption{Equations used to establish the P mass balance of the French sanitation system, and the datasets and parameters associated.}
	\label{fig:fig_P_mass_balance_equations}
\end{figure} 

\pagebreak

\subsection{Detailed methodology and discussion of uncertainties}\label{uncertainties}

A French report on extended measurement uncertainties on automatic sampling operations in wastewaters found that uncertainties for phosphorus were $<10\%$ \citep{AQUAREF2022Estimation}. That is our base rate uncertainties for flows whose estimation relies on direct measures of P. This concerns P in and out of WWTPs (flows A and B on Figure \ref{fig:fig_P_mass_balance_equations}) and industries discharge to sewers (flow H). In addition, there is a broader uncertainty for these flows, due to errors and perhaps incomplete reporting. Thus, we also observe there year-to-year variability to qualitatively discuss this uncertainty.
For the punctual parameters (P budget of individual autonomous systems, P excretions) we estimated the uncertainties by qualitative comparison to literature values.

\subsubsection{Uncertainty on phosphorus flows in and out of wastewater treatment plants}\label{A_P_uncertainties_flows}

\begin{figure}[h!]
\centering
    \includegraphics[width=0.95\linewidth, trim = {0 0 0 0}, clip]{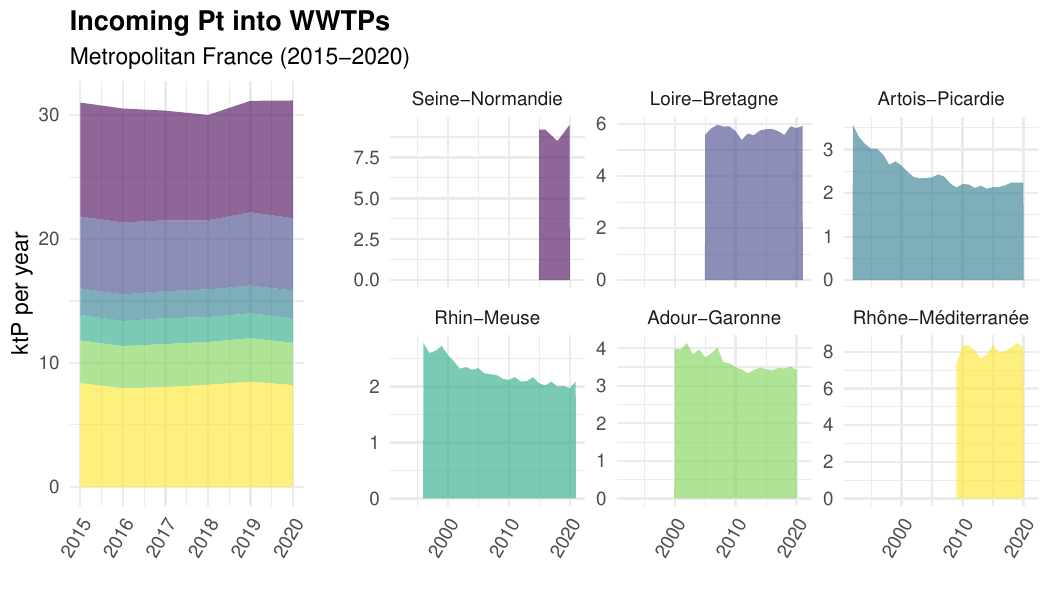}
    \caption{Reported P flows entering Waste Water Treatment Plants (WWTPs) at the national scale (left) and in the 6 water agencies, after data cleaning. For the national scale, only the reference period (2015-2020) is shown; for the 6 basins, all available data is shown. Actual flows may differ as reporting rates increase over time.}
	\label{fig:fig_P_flow_in}
\end{figure}

\begin{figure}[h!]
\centering
    \includegraphics[width=0.95\linewidth, trim = {0 0 0 0}, clip]{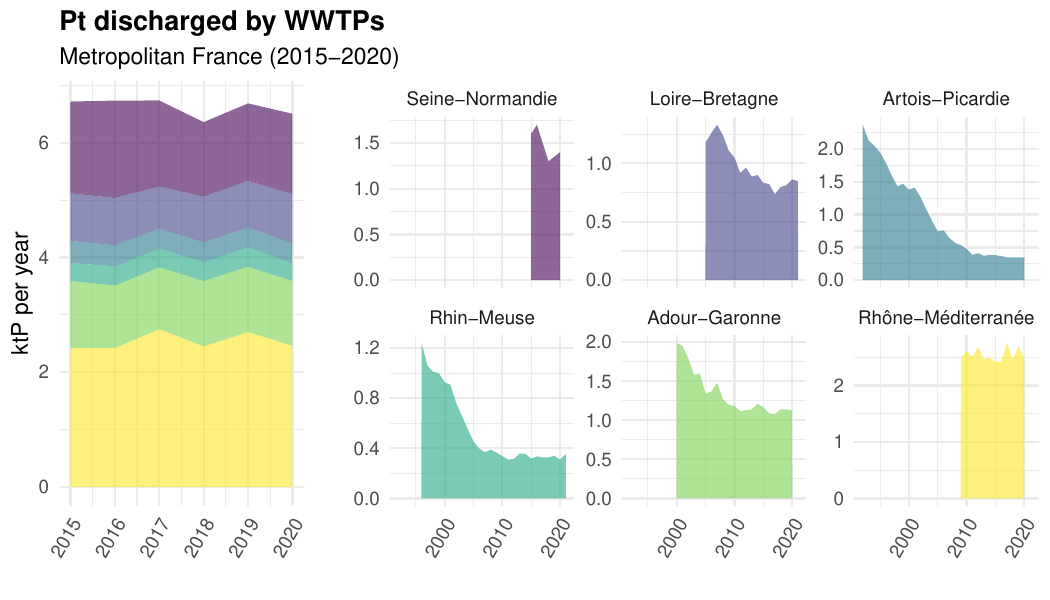}
    \caption{Reported P flows discharged by Waste Water Treatment Plants (WWTPs) at the national scale (left) and in the 6 water agencies, after data cleaning. For the national scale, only the reference period (2015-2020) is shown; for the 6 basins, all available data is shown. Actual flows may differ as reporting rates increase over time.}
	\label{fig:fig_P_flow_out}
\end{figure}

In the WWTP dataset used to quantify the P flows, the largest stations, with a capacity $>$100,000 population equivalent, are monitored at least every week, and often every day \citep{Légifrance2015Arrêté}. They represent $\sim$50\% of the total French WWTP capacity. Stations from 10,000 to 100,000 population equivalent, representing 30\% of the total capacity, are monitored at least every month. So, for 80\% of the national flow, there is a high certainty about the yearly-averaged reported value. After data cleaning, the year-to-year flows variability at the basin scale is small and on the order of 10\% maximum (Figure \ref{fig:fig_P_flow_in} et \ref{fig:fig_P_flow_out}). Considering this and the 10\% measure uncertainty on P flows in wastewaters, we chose a 10\% uncertainty for these flows (A and B on Figure \ref{fig:fig_P_mass_balance_equations})..

\subsubsection{Uncertainty on sludge production and destination}\label{A_P_uncertainty_sludge}

The sum of reported destination quantities  $r_t = \sum_i d_i$, about 0.9--0.95 Mt$\cdot$year$^{-1}$ of dry matter sludge in 2018-2021, is slightly lower than reported production quantities $p_t$ (about 1--1.1 Mt). Therefore, we extrapolate the total amount of sludge for each destination $i$ from the total production quantities, $d_i\frac{p_t}{r_t}$, considering that the fate of missing sludge in the destination report follows the same distribution as the reported quantities.

\begin{figure}[h!]
\centering
    \includegraphics[width=0.95\linewidth, trim = {0 0 0 0}, clip]{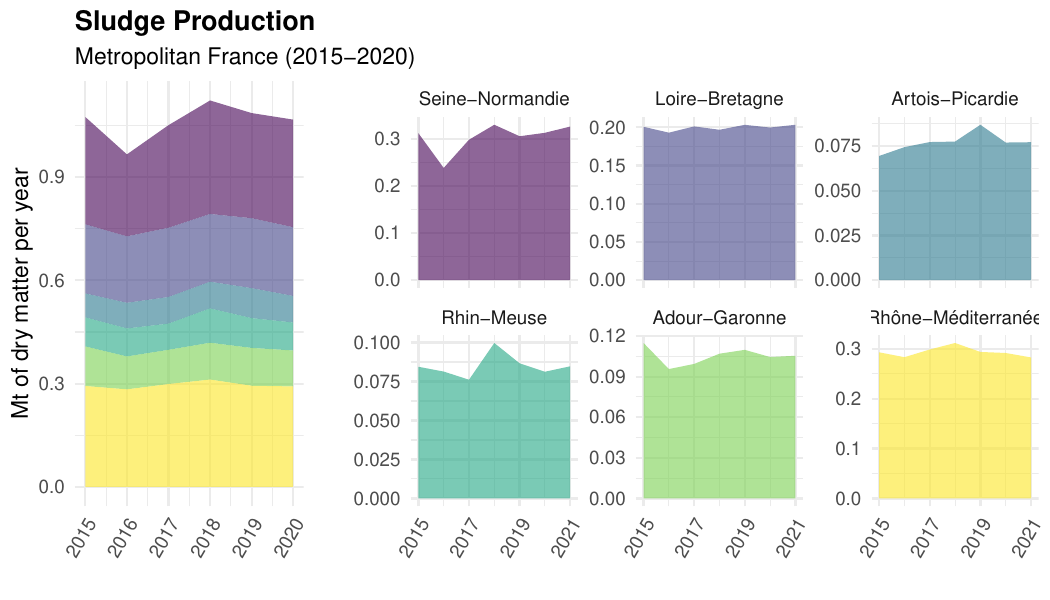}
    \caption{Reported Waste Water Treatment Plants (WWTPs) sludge production at the national scale (left) and in the 6 water agencies, after data cleaning. }
	\label{fig:fig_sludge_production}
\end{figure}

Following data correction, year-to-year variability in sludge production at the basin level is minimal, with a maximum fluctuation of $\sim$10\% (Figure \ref{fig:fig_sludge_production}). Similarly, the relative sludge destination has remained relatively stable since 2018 (45\% of sludge composted and 30\% directly spread on crops at the national scale) (Figure \ref{fig:fig_sludge_production}). Based on this, we use a 10\% uncertainty for sludge destinations.
Our results are comforted by the comparison to a French collective scientific expertise review on sludge composition \citep{Fuchs2014Effets}, whose reported P contents are similar to ours at the basin scale obtained by dividing P (C on Figure \ref{fig:fig_P_mass_balance_equations}) flow by reported sludge production. 

\begin{figure}[h!]
\centering
    \includegraphics[width=0.95\linewidth, trim = {0 0 0 0}, clip]{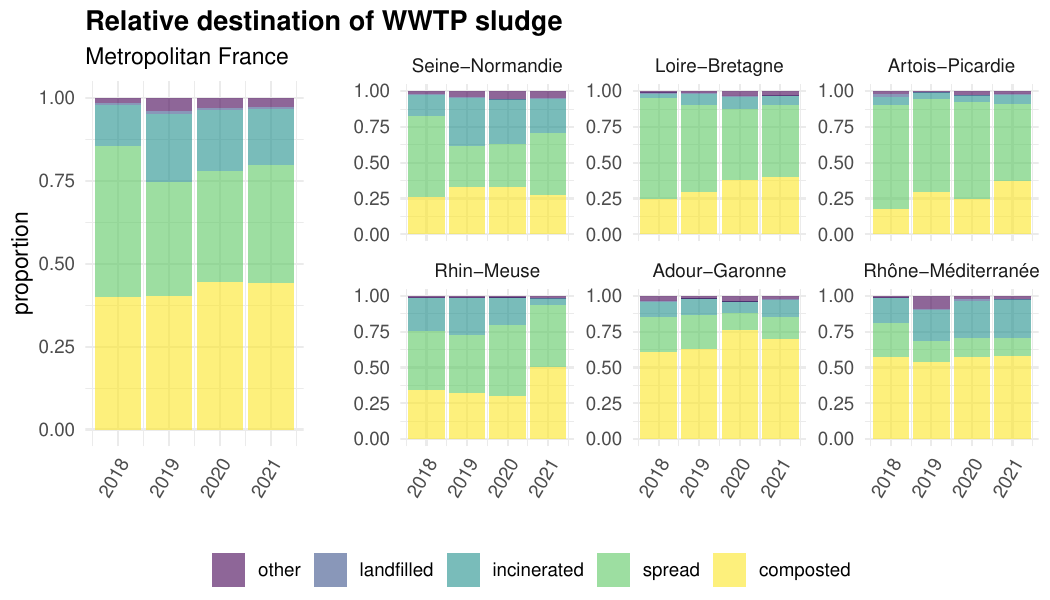}
    \caption{Reported Waste Water Treatment Plants (WWTPs) sludge relative destination at the national scale (left) and in the 6 water agencies, after data cleaning. Only the 2018-2021 period is shown, as previous years have very scarce and incoherent data. }
	\label{fig:fig_sludge_destination}
\end{figure} 

\begin{figure}[h!]
\centering
    \includegraphics[width=0.95\linewidth, trim = {0 0 0 0}, clip]{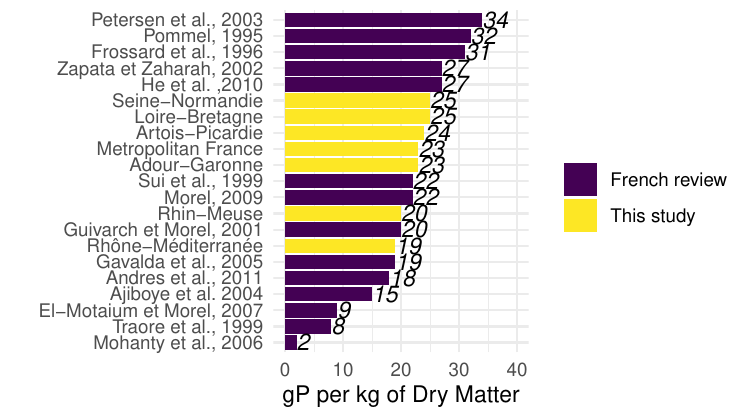}
    \caption{Comparison of sludge P content in Fuchs et al. (2014) and the value obtained from our P budget, at the national scale and in the 6 water agencies basins. }
	\label{fig:fig_compo}
\end{figure} 

\subsubsection{Large industries discharge in sewers}\label{A_P_industries}

\begin{figure}[h!]
\centering
    \includegraphics[width=0.95\linewidth, trim = {0 0 0 0}, clip]{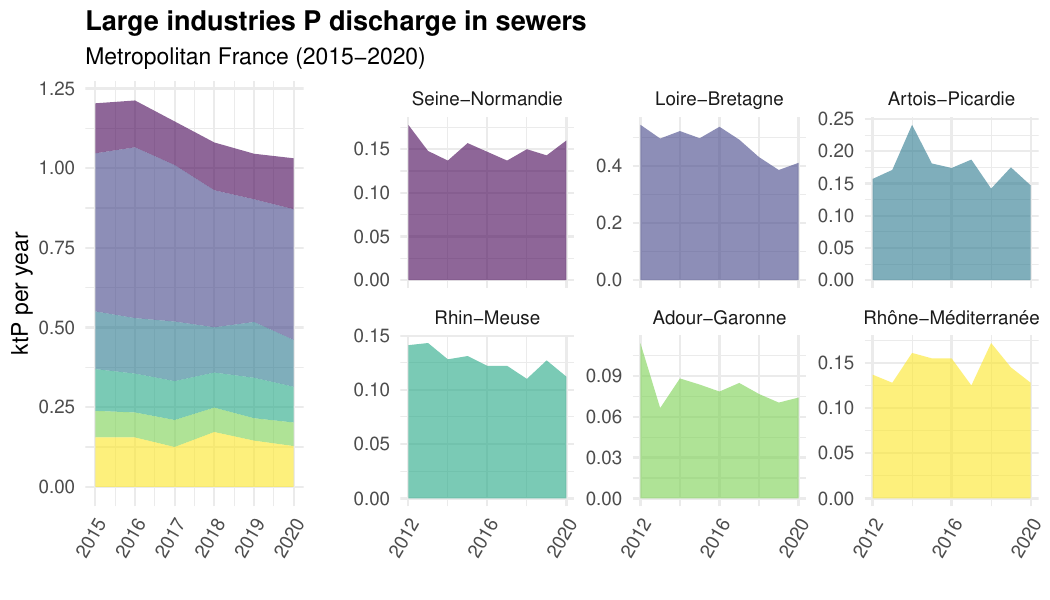}
    \caption{Reported P discharged by large industries at the national scale (left) and in the 6 water agencies, after data cleaning. }
	\label{fig:fig_P_industry}
\end{figure}

We chose to use the non-open GEREP database rather than the open-access “Géorisques” database \citep{ministere2023} that also reports industrial facilities discharging pollutants as part of the European Pollutant Release and Transfer Register (E-PRTR). “Géorisques” data only reports facilities discharging more than 5 tons of P per year \citep{Légifrance2007Décret}. On the other hand, the non-public GEREP data also includes facilities below this 5-ton threshold. While “Géorisques” only reports a few dozen facilities discharging P in sewers, the GEREP database reports about 1,200 facilities, which almost doubles the reported P flow. There is little year-to-year variability over the 2015--2020 period after data correction (figure \ref{fig:fig_P_industry}). Based on this little variability and the P flows measurement uncertainties of $<10\%$ \citep{AQUAREF2022Estimation}, we chose a 10\% uncertainty for this flow (H on Figure \ref{fig:fig_P_mass_balance_equations}). 

\subsubsection{Population and tourism}\label{A_P_population}

Between 2015 and 2020, the French population grew from 64.3 to 65.3 million residents. In the analysis, we used the population census from 2018, with 64.8 million residents, hence a maximum error smaller than 1\%.
Every year, France receives around 70 million tourists who come to visit the country, staying on average for a week. To this, one can add around 10 million tourists who transit through France from non-neighboring countries and around 100 million more who cross the border from a neighboring country, in both cases for a single day. Taken together, this represents an increase of around 9 days’ worth of excretion per capita over a year.
Overall, it means that we may underestimate the national input to the sewers from excreta by 3 \%.  However French people also travel abroad which leads to decreased local excretions. Precise assessment in the case of Paris Megacity \citep{Esculier2019biogeochemical} shows that outward and inward flows of people closely match. At the French level, the level of uncertainty of national inputs to the sanitation systems due to tourism is probably well below 3\%.

\subsubsection{Uncertainty regarding P excretion} \label{A_P_uncertainty_excretions}

\begin{figure}[h!]
\centering
    \includegraphics[width=0.95\linewidth, trim = {0 0 0 0}, clip]{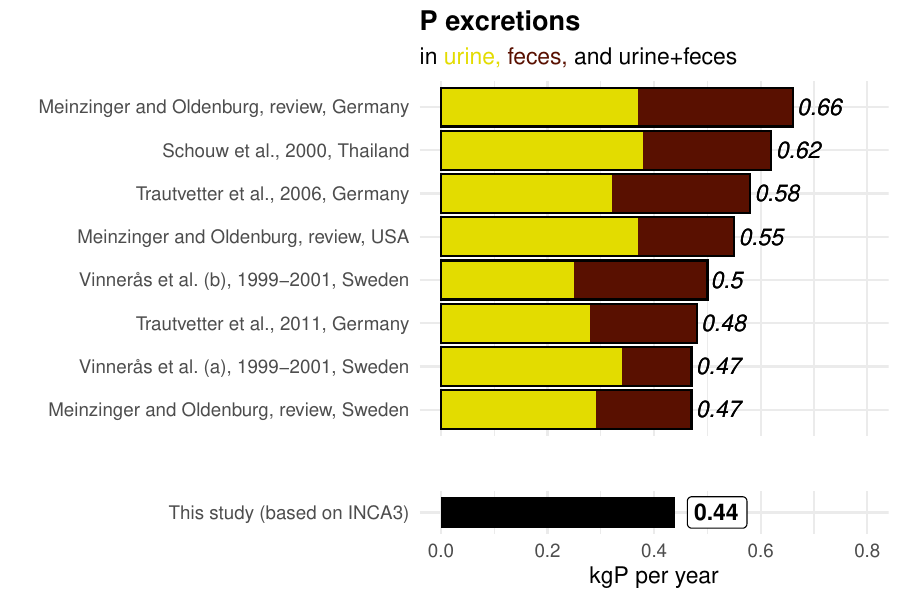}
    \caption{Comparison of actual measurements of P excretions in the literature to our study value.}
	\label{fig:fig_P_excretions}
\end{figure}

We assumed no P accumulation in adults' bodies and that all ingested P is excreted. The same approximation was made for children and adolescents, for whom \cite{Jönsson2004Guidelines} report a mean storage rate of 6\% between 2 and 17 years, which is consistent with medical data on bone growth \citep{Bass1999differing} and P intake from food. This is equivalent to 1.3\% over an 80-year lifespan, consistent with estimates from \cite{Scholz2014Sustainable} and higher than the 0.5\% from \cite{Liu2008Global}. Based on a mean body P content of 500--650 gP and daily intakes of 1--2 gP$\cdot$cap$^{-1}\cdot$day$^{-1}$ \citep{baccini2012metabolism, Elser2020Phosphorus:} over 80 years, the resulting annual population-level storage rate is also 1--2\%. 
Our estimate for the average P excretion at the national scale (0.44 kgP$\cdot$cap$^{-1}\cdot$year$^{-1}$, or 1.2 gP$\cdot$cap$^{-1}\cdot$day$^{-1}$) is consistent with the lower range of our literature review on P excretions (figure \ref{fig:fig_P_excretions}). This is exactly the same value as for Belgium in \cite{Papangelou2021Assessing}, also based on food intakes survey. Our method overlooks P body storage, which tends to overestimate P excretions by a few percent, but the food intake survey may underestimate actual P intakes, which could explain the average intake it is slightly lower than values reported in the medical literature that directly measure P in the excreta. Furthermore, standard deviation in medical data was high (typically 30\% of the mean) so 0.44 kgP$\cdot$cap$^{-1}\cdot$year$^{-1}$ is compatible with these values. 

Considering our review on Figure \ref{fig:fig_P_excretions}, we estimate that the estimation from Thailand is not representative of France, an industrialized country. Based the range of of values, we consider that the highest value for Germany (0.66 kgP$\cdot$cap$^{-1}\cdot$year$^{-1}$) is an outlier for our situation. We thus consider a plausible range of 0.44--0.68 kgP$\cdot$cap$^{-1}\cdot$year$^{-1}$. This constitutes our uncertainty range.

\subsubsection{Fate of P from Individual Autonomous Systems}\label{A_P_uncertainty_individual_system}

According to the Ministry of Ecological Transition, 15 to 20\% of the population is not connected to sewers and relies on Independent Wastewater Treatment \citep{ministere2023}. We thus use these figures for our uncertainty range of 15\%-20\%. We considered that for people in individual autonomous systems, 20\% of their excretions happen outside their home, in public spaces connected to sewers. We choose and uncertainty range associated with this parameter of 5-35\%.
5\% would correspond to elder people who rarely leave their house while the 35\% corresponds to people who work (210 days on average, in France) in a workplace connected to sewers and excrete on average 60\% of their phosphorus flow during their time at work.

We did not consider excretions in diapers for newborns and elderly people, which do not enter sewers nor individual autonomous systems but are directly landfilled or incinerated.

For the P mass balance of individual autonomous systems, LCA analyses of a septic tank + sand filter from \citep{Risch2021Applying} and a French report \citep{Catel2017Inventaires} were used. The P destination for \cite{Risch2021Applying} is 83\% in underground and 17\% in sludge, while the French report gives slightly different figures of 90\% and 10\% respectively. We assume an intermediate repartition of 85\%--15\% and that all sludge is directed to wastewater treatment plants. This is the figure used in \citep{Linderholm2012Phosphorus}, but lower than the 75\%--25\% used by \cite{Senthilkumar2014Phosphorus}. Based on this, we estimate the uncertainty associated with the different P destinations to be in the range of 10\% to 20\% for sludge and 80\% to 90\% for underground.

\subsubsection{Phosphorus loss before WWTPs}\label{A_P_loss_before_WWTP}

Part of the P entering sewers does not reach the WWTP entrance, due to combined sewers overflows and sewers leaks. Data are non-existent concerning this latter point, and very scarce for overflows. Adour--Garonne and Loire--Bretagne basins report data from a sample of the largest WWTPs, where direct discharges represent respectively 7\% and 15\% of the flows finally entering the WWTPs. For these 2 basins, we extrapolate the loss rate of this WWTPs sample to the whole basin. No data was available for the Rhône--Méditerranée basin, so we used the loss rate of the other French Southern basin, Adour--Garonne. For the 3 other basins (Rhin--Meuse, Seine--Normandie and Artois--Picardie), we use estimates from their respective experts’ assessment Etat des lieux (respectively 20\%, 10\% and 20\%). The resulting national mean (weighted by incoming P flow quantity) is 10\%. Based on the diverging values among the different basins, we estimate the actual value to be between 5\% and 20\%, which is used as our uncertainty range. For comparison, \cite{Scholz2014Sustainable} estimate wastewater losses due to leakage before reaching WWTPs in the EU15 at 5--10\%.

\pagebreak

\subsection{Nutrient flows uncertainties}
\label{app:uncertainty-prop}

\begin{table}[h!]
    \centering
\caption{Parameters values and their associated uncertainties to compute flows detailed on Figure \ref{fig:fig_P_mass_balance_equations}. The "value" column reports our parameter best estimate. For point estimates of single parameters, we consider a uniform distribution with a range of values, based on our discussion in the previous sections. For measured flows with multi-year values, we consider a 10\% uncertainty, based on reported measurement uncertainties \citep{AQUAREF2022Estimation} and year-to-year variability. We consider that the $\pm10\%$ range contains 95\% of the values, i.e. $2\sigma/\mu = 10\%$}
\label{tab:uncertainties_parameters}
    \begin{tabular}{>{\raggedright\arraybackslash}p{0.34\linewidth}>{\centering\arraybackslash}p{0.14\linewidth}>{\centering\arraybackslash}p{0.08\linewidth}>{\centering\arraybackslash}p{0.12\linewidth}>{\centering\arraybackslash}p{0.2\linewidth}}
        \toprule
         \multirow{3}{=}{\textbf{parameter}} &  \textbf{used in flow on Figure \ref{fig:fig_P_mass_balance_equations}}&    \multirow{3}{=}{\textbf{value}} & \multirow{3}{=}{\textbf{distribution}}&\textbf{range for uniform or $\boldsymbol{2\sigma/\mu}$ for normal}\\
    		\hline
         ktP coming to WWTPs&  A&    30.8&normal&10\%\\
         ktP discharged from WWTPs&  B&    6.6&normal&10\%\\
         \% sludge composted or spread&  D&    77\%&normal&10\%\\
         \% sewers discharge&  F, G&    10\%&uniform&5--20\%\\
 ktP industries discharge to sewers& H&  1.1&normal&10\%\\
  kgP/cap in human excretions& I&  0.44&uniform&0.44--0.58\\
 \% people in IAS& J, M&  15\%&uniform&15--20\%\\
 \% excretions outside home& J&  20\%&uniform&5--35\%\\
 \% IAS to diffuse loss& N&  85\%&uniform&80--90\%\\
 \% IAS to WWTP& O&  15\%&uniform&10--20\%\\
    		\bottomrule
    \end{tabular}

\end{table}

\begin{table}[h!]
    \centering
\caption{Flows best estimate values and uncertainties. Best estimate is simply obtained by applying equations reported in Figure \ref{fig:fig_P_mass_balance_equations} to the values reported in \ref{tab:uncertainties_parameters}. Uncertainties are obtained by applying a Monte-Carlo Procedure to the distributions reported \ref{tab:uncertainties_parameters}.}
\label{tab:uncertainties_results}
    \begin{tabular}{>{\raggedright\arraybackslash}p{0.42\linewidth}>{\centering\arraybackslash}p{0.1\linewidth}>{\centering\arraybackslash}p{0.2\linewidth}>{\centering\arraybackslash}p{0.18\linewidth}}
        \toprule
         \multirow{2}{=}{\textbf{Flow description}} &  \textbf{Letter in Fig \ref{fig:fig_P_mass_balance_equations}}& \textbf{Best estimate (ktP/year)} & \multirow{2}{=}{\centering\textbf{95\% CI}}\\
    		\hline
 P entering WWTPs from sewers& A&30.8&27.8--33.8\\
 P discharged to waters by WWTPs& B&6.6&6--7.2\\
 P trapped in sludge& C&24.9&21.9--28.3\\
 Sludge recycled to agriculture& D&19.2&16.3--22.5\\
 Sludge landfilled or incinerated& E& 5.7&3.8--7.9\\
 P entering sewers& F& 34.2&31--40\\
 P lost in sewers or stormwaters& G& 3.4&1.8--7.6\\
 P discharged by large industries to sewers& H& 1.1&1--1.2\\
 P excreted by French population& I& 28.4&28.4--37.2\\
 Excretions going to IAS& J& 3.5&3.3--6.1\\
 Excretions going to sewers& K& 24.9&24.3--32.4\\
 Residual P in sewers& L& 8.1&0--12.5\\
         Residual P to IAS&  M&  1.5&0--2.8\\
         P not trapped in IAS&  N&  4.2&3.5--6.8\\
         P trapped in IAS sludge sent to WWTP&  O&  0.7&0.5--1.4\\
    		\bottomrule
    \end{tabular}

\end{table}

\pagebreak

\subsection{P footprint and contribution to self-sufficiency calculation}\label{P_footprint}

We compute the P footprint of food supply (before food waste) for the world and France based on \citet{Sutton2013Our} and \citet{Esculier2019biogeochemical}. P footprint, P food supply and P use efficiency (PUE) are linked by the relation $\text{P}_\text{supply} = \text{PUE}\cdot\text{P}_\text{footprint}$, $\text{P}_\text{supply}$ being the phosphorus contained in the food supply. P footprint and PUE are computed for: diet = animal+plant products together; animal products alone; plant-based products alone. All figures and calculation are detailed in our supplementary spreadsheet.

In \citet{Sutton2013Our}, we use the numbers in pages 23--24 and particularly Figure 3.2. We use reported animal and plant $\text{P}_\text{supply}$. We combine this with reported PUE for crops $\text{PUE}_\text{c}$ and animal products $\text{PUE}_\text{animal, ingestion}$ to compute the P footprints. For animal products, the reported $\text{PUE}_\text{animal, ingestion}$ only concerns the efficiency starting at animal ingestion. To account for the full-chain animal product PUE ($\text{PUE}_\text{a}$), we have to also consider the PUE of crops fed to animals. We consider 2 extreme cases. First, if all forage given to animals come from unfertilized grasslands, then forage PUE is 100\% and $\text{PUE}_\text{a} = \text{PUE}_\text{animal, ingestion}\cdot 100\%$. The other case is if all feed given to animals comes from crops. Then $\text{PUE}_\text{a} = \text{PUE}_\text{animal, ingestion}\cdot\text{PUE}_\text{c}$. This gives a range of animal P footprint, computed by  $\text{P}_\text{footprint} = \text{P}_\text{supply}/\text{PUE}$, with the real footprint being somewhere in between these extremes.

For \citet{Esculier2019biogeochemical}, we use numbers in Figure 5 (for PUE, supply and footprint) and Figure 8 (for P ingestion/excretion). The previous issue of feed PUE given to animal does not apply here since it is already computed. The animal P footprint, supply and PUE relate to the combination of ``intensive livestock farming'' and ``mixed crop and livestock farming''. Contrary to \citet{Sutton2013Our}, P footprint numbers are already reported, and we compute PUE from $\text{PUE} = \text{P}_\text{supply}/\text{P}_\text{footprint}$. The case of ``other territories'', where only P supply is reported but not P footprint, is extrapolated from the PUE of their particular farming system (crop farming; mixed crop and livestock farming; intensive livestock farming).

In both cases, the total P food supply is the sum of P content of animal-based and plant-based products, but also fish and additives. This P supply is different than the P ingested/excreted, because food waste happens in between.

The potential share of supply $S_\text{P,excr}$ that could be covered by recycling P in excretions (not prioritizing plant-based products) is obtained through $S_\text{P,excr} = \text{P}_\text{excretions}\cdot\text{PUE}_\text{diet}/\text{P}_\text{supply}$. For the case when P excretions are prioritized to plant-based products, we first compute $\text{P}_\text{crops}^\text{potential} = \text{P}_\text{excretions}\cdot\text{PUE}_\text{c}$. If this quantity exceeds the P supply of plant-based products, we allocate the surplus to animal products $\text{P}_\text{remaining}\cdot\text{PUE}_\text{a} = (\text{P}_\text{crops}^\text{potential} - \text{P}_\text{crops})\cdot\text{PUE}_\text{a}$, such that:
\begin{equation*}
S_\text{P,excr}^\text{plant prioritized} = (\min(\text{P}_\text{crops}, \text{P}_\text{excretions}\cdot\text{PUE}_\text{c}) + \Theta(\text{P}_\text{remaining})\cdot\text{PUE}_\text{a})/\text{P}_\text{supply}
\end{equation*}
where $\Theta(x) = x$ if $x > 0$ and $\Theta(x) = 0$ otherwise.

\pagebreak

\begin{table}[h!]
    \caption{Phosphorus footprint from plant and animal products in France and in the world used to compute the share of P in food supply that could be covered by P in excreta. See supplementary spreadsheet for detailed computations.}
    \label{suptable_P_footprint}
    \centering
        \begin{tabular}{m{0.15\linewidth} M{0.3\linewidth} M{0.3\linewidth}}
        \toprule
                & \shortstack{\textbf{World (MtP)}\\(years 2000-2010)\\ \cite{Sutton2013Our}} & \shortstack{\textbf{Paris Megacity (ktP)}\\(year 2012)\\ \cite{Esculier2019biogeochemical}} \\ 
            \hline
                 P in excretions / ingestions & 4 & 4.3 \\ 
            \hline
                \multirow{5}{=}{P in domestic food supply} & Plant 2.9 & Plant 2.2 \\
                 & Animal 1.5 & Animal 3.9 \\
                 & Fish 0.2 & Fish 0.5 \\
                 & Additives 0.5 & -\\
                 & Total 5.1 & Total 6.6\\
            \hline
                \multirow{3}{=}{P footprint (excluding fish\\and additives)} & Plant 5.8 & Plant 2.1 \\
                 & Animal 15--30 & Animal 53.2 \\
                 & Plant + Animal 20.8--35.8 & Plant + Animal 55.3 \\
            \hline
                \multirow{3}{=}{P use efficiency (supply/footprint)} & Crop $50\%$ & Crop $105\%$ (\textit{mining})\\
                 & Animal 5--10\% & Animal $7\%$\\
                 & Plant + Animal 12--21\% & Plant + Animal $11\%$\\
            \bottomrule
        \end{tabular}
\end{table}

\pagebreak

\subsection{Data sources used for the P budget.}\label{tab:data_sources}

\begin{table}[h!]
    \caption{Data sources used for the P budget.}
    \label{table_P_data_sources}
    \centering
    \begin{tabular}{m{0.15\linewidth}m{0.25\linewidth}m{0.40\linewidth}}
        \toprule
            \multicolumn{2}{c}{\textbf{data}} & \textbf{sources} \\ 
        \midrule
            \multicolumn{2}{l}{P discharge to sewers from each individual industry} & GEREP database, nonpublic extension of the open access \href{https://www.georisques.gouv.fr/installations-industrielles-rejetant-des-polluants}{georisque French database}, part of the European Pollutant Release and Transfer Register (E-PRTR). \\
            \multicolumn{2}{l}{Population by age and sex for each French city.} &  \cite{INSEE2022Population} \\
            \multicolumn{2}{l}{Sludge production and destination for each WWTP.} & \cite{PortailAssainissementCollectif2023} \\
        \hline
            \multirow{15}{=}{P flows in and out of each individual wastewater treatment plants, for each water agency basin} & Artois Picardie basin \newline years 1992-2021 & Artois-Picardie agency \href{https://www.artois-picardie.eaufrance.fr/cartes-et-donnees/les-donnees-sur-l-eau-du-bassin-artois-picardie/}{website} \\
         & Rhin-Meuse basin,  \newline years 1996-2021 & Rhin-Meuse agency \href{https://rhin-meuse.eaufrance.fr/telechargement?lang=fr}{website}  \\
         & Seine-Normandie basin \newline years 2015, 2016, 2018, 2020 & Etat des lieux (status reports) data, only for 2015, 2016, 2018, 2020 communicated through mail. We also have the SIAAP data over 2007-2021, consisting of the 6 largest WWTP representing 50\% of the basin nutrient flows. \\ 
         & Loire-Bretagne basin \newline years 2005-2021 & Communicated by email, publicly shareable \\
         & Adour-Garonne basin \newline years 2000-2020 & Adour-Garonne agency website, \href{http://adour-garonne.eaufrance.fr/catalogue/581d5f70-558c-49e4-8d77-5bd4fe974b62}{link} for nutrient flows discharge and \href{http://adour-garonne.eaufrance.fr/catalogue/42f43670-099d-11de-97dd-001517506978}{link} for WWTP description  \\
         & Rhône-Méditerranée basin \newline years 2009-2020  & Rhône-Méditerranée agency \href{https://www.rhone-mediterranee.eaufrance.fr/telechargements/bibliotheque-de-telechargement-de-donnees-sur-leau}{website}  \\
        \bottomrule
    \end{tabular}
\end{table}

\pagebreak

\printbibliography

\end{document}